
\input phyzzx

\def\ra{\rightarrow}
\def\prd#1{{\sl Phys.~Rev.}~{\bf D#1}\ }

\def\plett#1{{\sl Phys.~Lett.}~{\bf #1B}\ }
\def\nphys#1{{\sl Nucl.~Phys.}~{\bf B#1}\ }
\def\abstract{\vskip\frontpageskip\centerline{\twelverm ABSTRACT}
              \vskip\headskip }
\def\NEU{\address{Department of Physics\break
      Northeastern University, Boston, Massachusetts 02115}}
\def\us#1{\undertext{#1}}
\def\quarter{\ifmmode{\textstyle{1\over 4}}\else ${\textstyle {1\over 4}}$\fi}
\def\half{\ifmmode{\textstyle{1\over 2}}\else ${\textstyle {1\over 2}}$\fi}

\REF\corn{J.~M.~Cornwall, \plett{243}  (1990) 271.}
\REF\hgs{H.~Goldberg, \plett{246} (1990) 445.}
\REF\hgv{H.~Goldberg, ``Non-Perturbative Multiparticle Tree Graphs in
the Standard Electroweak Theory'', Northeastern University preprint
{\sl NUB-3029/91-Th}, September 1991, to be published in {\sl Physical
Review D.}}
\REF\matt{P.~B.~Arnold and M.~P.~Mattis, \prd{42} (1990) 1738;
L.~McLerran, A.~Vainshtein, and M.~Voloshin, {\it ibid} 180;
S.~Yu.~Khlebnikov, V.~A.~Rubakov, and P.~G.~Tinyakov, \nphys {350}
(1990) 441.}
\REF\voloshin{M.~B.~Voloshin, \prd{43} (1991) 1726. See also a recent
discussion by J.~M.~Cornwall and G.~Tiktopoulos, UCLA preprint
UCLA/91/TEP/55.}
\REF\bachas{C.~Bachas, Ecole Polytechnique preprint A089.1191,
December 1991.}
\REF\zakh{V.~I.~Zakharov, Max Planck preprint MPI-PAE-PTH-11-91, March
1991.}
\REF\shifman{M.~Maggiore and M.~Shifman, Minnesota preprint
TPI-MINN-91-17-T-REV, February 1992.}
\REF\hgn{H.~Goldberg, Northeastern preprint NUB-3044/92-TH, to be published.}
\def\one8{1\Rightarrow 8}
\def\aemax{|A_8|_{\rm max}}
\def\onen{1\rightarrow N}
\def\a1n{A_{\onen}}

\def\calp{{\cal P}}
\def\g2{g^2}
%
\Pubnum{\sl NUB-3043/92-TH}
\date{\sl March~1992}
\pubtype{}
\titlepage
\hoffset 0.36in
\voffset 0.30in
\title{\bf Exact Nonperturbative Unitary Amplitudes for $\one8$
Transitions in a Field Theoretic Model }
\vskip .8cm
\author{H. Goldberg and M.~T.~Vaughn}
\vskip .5cm
\NEU
\vskip .4cm

\abstract

We present a quantum mechanical model with an infinite number of
(discrete) degrees of freedom, which can serve as a laboratory for
multiparticle production in a collision. There is a cubic coupling
between modes without, however, any problems associated with unstable
ground states. The model is amenable to precise numerical calculations
of nonperturbative $\onen$ transition amplitudes. On an ordinary
workstation, time and memory limitations effectively restrict $N$ to
be $\le\ 8,$ and we present results for this case. We find (1) that there is
reasonable period of time for which there is a constant rate for the
$\one8$ transition; (2) at the end of the linear period, the eight
particle amplitude attains a maximum value $\aemax$ which is about
$3-4$ orders of magnitude larger than the comparable amplitude for
excitation of the $N=8$ state in the anharmonic oscillator;
(3) for values of the coupling in the region where the Born
approximation fails, the amplitude
is much larger than the naive estimates $A_8\simeq
\exp{(-1/\g2)}\ $ or $\ \exp{(-8)};$ it is more like
$A_8\sim\exp{(-0.20/\g2)}.$

\vskip 1.5cm
\endpage

\hoffset 0.37in
\voffset 0.31in
\frontpagefalse
\pagenumber=1

The calculation of tree-level amplitudes for the production of $N$
bosons at fixed momenta indicates a failure of perturbation theory
when $N$ is large. For a quartic coupling $\quarter \g2\phi^4,$ the
$\onen$ amplitude $\a1n\sim g^{N-1}\ N!$ for large $N.$ [\corn,\hgs]
Thus, perturbation theory may become untrustworthy when $N\sim 1/g.$
The origin of the problem is simply the overcoming of the $g^N$
suppresion by the enhancement factor $N!$ which originates in the
coherent addition of the $O(N!)$ graphs obtained by permutation of the
momenta. The coherence has been established in the non-relativistic
region both in the scalar field theory [\hgs] and in the standard
model, including the production of $W$'s and $Z$'s [\hgv]. For theories
with a continuous momentum spectrum in an infinite volume, the
transition rate for large $E, N,\ $ fixed $E/N$ is $\Gamma\sim
|\a1n|^2/N!,$ which violates unitarity for $N\gsim 1/\g2.$ Of course,
it is clear that for $N$ somewhere in this range, loops will
contribute equally to the tree graph: each tree may generate $O(N^2)$
final state scatterings, at a cost of $\g2$ for each. Depending on the
coherence of the loop corrections, these may become unsuppressed when
$N\sim 1/g$ or $N\sim 1/\g2.$

Subsequent to these studies, there have been several papers written
which argue for an exponential suppression of the amplitude. That is,
if we define the `holy grail' function $F$ [\matt] via
$$\a1n\sim e^{F/\g2}\ \ \ ,\eqn\fdef$$
then exponential suppression means that for $N\gsim 1/\g2,$
$$F(N,\g2)\sim O(-1)\qquad {\rm or}\quad O(-N\g2)\ \ .\eqn\fsim$$
One approach has been based on finding a bound for the $\ket{{\sl
ground\ state}}\ra \ket{N}$ transition for the anharmonic well  in
Schr\"odinger mechanics; either a calculation based on the $WKB$
approximation [\voloshin] or a variational study [\bachas] shows that
the overlap matrix elements $\VEV{N|x^2|0}$ (or $\VEV{N|x|0})$ are
exponentially suppressed, with an $F$ which is negative and
monotonically decreasing as a function of the variable $\lambda=N\g2.$
$(\g2$ is the coefficient of the quartic coupling.) Thus, for fixed
$N$ (which we will be constained to work with), $F$ in the quantum
mechanical model is a decreasing
function of $\g2.$ In a second approach [\zakh,\shifman], a bound on
$\a1n$ is claimed to exist on the basis of a Lipatov-type analysis.
Here again, the bound is stated to be exponential: $F\sim O(-1)$ for
$N\gsim 1/\g2.$

In our view, there are uncertainties in the application of both these
approaches to the problem at hand. In the first, we have no argument
with the bound in terms of quantum mechanics {\it per se.} However,
since the physics of the q.m. problem involves the excitation of the
single $N$'th level, while the multiparticle process addresses the
production of $N$ different low-energy quanta (for fixed momenta),
some hesitancy may be justified before adopting for the
multiparticle problem the exponential suppression from the q.m. bound.
With respect to the Lipatov-type analysis, at least in the approach of
Ref. [\shifman], we have found ourselves uncertain in extrapolating
the results from fixed numbers of legs to a large number $N\sim
1/\g2;$ without being critical, we wish to observe the suppression
(should it exist) in a much more direct manner; that is, we would
really like to see how  unitarity for $\a1n$ gets restored through
rescattering. If there is exponential suppression, we would like to
find the actual value of $F.$

In this work, we propose a simple field-theoretic model in which
exact unitary amplitudes for transitions from 1 to $N$ particles can
be calculated. In principle, these can be done exactly analytically;
in practice, we will resort to a numerical calculation.  Even
numerically, we cannot go much beyond a $\one8$ transition; and even
this process will  require the solution of 1336 coupled first order
differential equations. Nevertheless, the resulting amplitude will be
an exact, unitary, nonperturbative solution of coupled
time-dependent Schr\"odinger equations; and the Born approximation
will exhibit (for $N=8)$ the $g^{N-1}\ N!$ blowup of the scalar tree.

   The model may be conveniently defined through its Hamiltonian
$$H= \sum_{k=1}^\infty a^{\dagger}_k a_k + \half g\ \calp\ \sum_{j,k=1}^\infty
a^{\dagger}_{j+k}\ a_j\ a_k\ \calp + \ \ h.c.\ \ \ \  .\eqn\ham$$
The modes labeled by $i,j,k$ will be called ``momenta'', and the action of the
hermitean projection operator $\calp$ on a state vector $\ket{\psi}$ is as
follows:
$$\calp\ket{\psi}=0 $$
if there is more than one quantum in the state with any given momentum
$$\calp\ket{\psi}=\ket{\psi}$$
otherwise.

$\calp$ has been introduced in order to mimic the infinite volume effect of
field theory (in box normalization): namely, we do not generally concern
ourselves with amplitudes for transitions to states with more than one particle
in a given (discrete) momentum state. Thus, we exclude ``laser''
effects, with their attendant $\sqrt{n_i!}\ $ factors in matrix elements.
In
this sense, these are ``hard core'' bosons: there
is a fermionic exclusion principle without imposing
anticommutation relations and antisymmetrization.
Other than that, $H$ resembles a $\phi^3$ field theory in a cavity, with no
momentum dependence to the unperturbed energies. It is also a kind of matrix
model.

It is an important consequence of \ham \ that the momentum operator
$$P=\sum_{k=1}^{\infty}\ k\ a^{\dagger}_ka_k\ \ \ \eqn\mom $$
is a constant of the motion. Thus, the Hilbert space factorizes into subspaces
with definite $P.$ Because of the positivity of all of the momenta, {\it these
will be finite dimensional subspaces.} In considering the subspaces with
definite $P,$ we may conveniently think in terms of a {\us{maximal}} state,
namely an $N$-particle state with momenta $k=1,2,\ldots N.$ This state will
have a total momentum $P=N(N+1)/2.$ We will our study to such sectors, labeled
by $N,$ which have this maximal state in their spectrum.

In general, the total number of states for a given $P$ can be obtained as the
exponent of $x^P$ in the expansion of the generating function
$$\prod_{j=1}^\infty (1+x^j)= \sum_{P=0}^\infty {\cal N}_P\
x^P\ \ \ .\eqn\gen$$

If we consider the symmetric trees corresponding to $N=2,4,8,16\ldots$
$(P=3,10,36,136\ldots),$ we find that the number of states to be considered are
2, 10, 668, 7 215 644,... respectively. Even  going from $N=8$ to $N=9$ means
increasing the subspace from 668 to 2048 states. Thus, in a practical decision
which is based on CPU time considerations, we restrict
the present study to the case $N=8.$

For a given $N,$ there will be states with $n=1,2,\ldots N$ particles. The
cubic interaction $gV$ will couple a state $\ket{n,r}$ to states $\ket{n+1,s}$
and $\ket{n-1,s}$, where $r,s$ label the different states with a given number
of particles. For example, with
$N=8\ (P=36),$ $V$ will couple the 4-particle state
$\ket{1,3,8,24}$ to various 5-particle states (such as $\ket{1,3,8,11,13}$ but
not $\ket{1,3,8,12,12})$ and 3-particle states (such as $\ket{1,11,24}).$

The energy difference between (unperturbed) states coupled by $V$ is always
$\pm 1.$ Thus, if we work in the interaction representation, so that the the
amplitude to be in the (unperturbed) state $\ket{n,r}$ at time $t$ is
$A_n^r(t),$ then the Schr\"odinger dynamics is governed by the coupled
equations, for $n=1,2\ldots N$ and $A_0=A_{N+1}=0:$
$$\eqalign{  i{\dot A}_n^r(t)\ =\ & g\ \left( \sum_s\VEV{n,r|V|n-1,s}
\ e^{it}\ A^s_{n-1}(t)\right.\crr
&\ +\ \left.\sum_s\VEV{n,r|V|n+1,s}\ e^{-it}\ A^s_{n+1}(t)\right)
\ \ \  .\crr}\eqn\schr$$

For the case $N=8,$ we have calculated explicitly
the number of states for each $n,$ their momentum content, and the
$668\times 668$ transition matrix (whose elements are $g$ or 0). For example,
the number of states for $n=1,\ldots 8$ are (1, 17, 91, 206, 221, 110,
21, 1)
respectively. We have solved these equations subject to the boundary condition
$$A_1(0)=1,\qquad A_n^r(0)=0\ \ ,\ n\ne 1\ .\eqn\bc$$
The {\us{Born (tree) approximation}} is obtained from \schr \
by keeping only the
coupling of $n$ to $n-1$ (no rescattering). This set of equations can be solved
analytically. For the amplitude $A_8$ (there is only one), we obtain
$$A_8 = 427,206\ {g^7\over 7!}\ \left(1-e^{it}\right)^7
\ \  ,\eqn\bornt$$
where the first numerical factor is the result of successive matrix
multiplications leading from the single one-particle state $\ket{36}$ to the
single 8-particle state  $\ket{1\ldots 8}.$ It is an ``entropy'' factor, since
this multiplication counts the the number of paths between these states.
{}From \bornt\ we find
$$|A_8^{\rm Born}|_{\rm max}\simeq 0.27\ g^7\ 8!\ \  ,\eqn\bornmax$$
a relation in accordance with the tree level result in the $\phi^3$ theory
[\hgs]. We note that the Born amplitude will exceed the unitary bound when
$\g2>0.07:$ this is then an upper limit on $\g2$ for weak coupling in the case
of $N=8.$

\noindent{\us{Numerical Results}}. In integrating the coupled 1336 (real)
Eqs.\ \schr , we have used an adaptive Runge-Kutta routine, and have found that
unitarity is satisfied to better than 1 part in $10^7$ for all times relevant
to the discussion.  In Fig. 1 we show the probability $|A_8(t)|^2$ {\it vs.}
$t$ for the ``critical'' coupling $\g2=0.07.$ Note that $|A_8|^2$ has a maximum
value of about $3\times 10^{-4},$ and enjoys a linear rise (constant transition
rate) for a reasonable amount of time. We find for all $\g2$'s of interest,
$$\left({d\over dt}|A_8(t)|^2 \right)_{\rm max}
\sim |A_8(t)|^2_{\rm max}\ \ , \eqn\pdot$$
with the unit of time set by the basic frequency,$\ \omega=1.$
Thus, for ease of
comparison with the quantum mechanical results, we will phrase the discussion
in terms of $|A_8(t)|_{\rm max},$ and compare this with the
variational bound to the transition matrix
element $\VEV{8|x|0}$ of ref. [\bachas]. For the sake of general
interest,
we  show in Fig.~2 \ some of the long time behavior of the probability
of staying in the original state. It can be seen that even in the
nonperturbative regime, there is a residual quasi-periodicity which we
expect will dissipate over very long times.

In Fig. 3, we present a plot of $|A_8|_{\rm max}$ {\it vs.} $\g2.$ For
comparison, the dotted line shows the variational
upper bound on the amplitude
$\VEV{8|x|0}$ in ref. [\bachas] for the anharmonic oscillator with
potential $\half x^2 + \quarter \g2x^4.$ Without particular
justification, we have identified this $\g2$ with ours. In the weak
coupling regime, there are $3-4$ orders of magnitude difference
between these amplitudes. It should be noted that both our 8-particle
state $\ket{1,2,3,4,5,6,7,8}$ and the oscillator state $\ket{8}$ are
normalized to unity. The
difference between the dynamics becomes more apparent when the $\g2$
behavior of the amplitudes (for fixed $N=8)$ is compared. In Fig.~4,
we have plotted the holy grail function $F$ {\it vs.} $\g2$, with the
variational upper limit from the anharmonic oscillator [\bachas] (for
fixed $N=8)$ shown for comparison. Again, one may note the great
difference in behavior of the two amplitudes. We may also note that
the Lipatov-type analysis might predict $F\sim O(-1)$ for $\g2\simeq
0.05-0.10,$ the region where the tree approximation breaks down. We
find $F\simeq -0.20$ in that region. While it is difficult to claim
that this indicates a lack of exponential suppression, it is
nevertheless a possible hint that any suppression, should it exist, is
weaker than believed. In a subsequent work [\hgn] in which an
extension to arbitrary $N$ is presented, it will be seen that
exponential suppression is extremely weak $(|F|\ll 1)$
in the weak coupling limit.

\noindent{\us{Conclusions}}. We have proposed a model, reminiscent of
$\phi^3$ field theory, in which  $\onen$ transition amplitudes can be
calculated in principle. CPU time and disc storage limitiations have
limited the present considerations to a maximum of $N=8.$ In contrast
to quantum mechanical calculations in the anharmonic oscillator model,
the ``$N$'' in the present case refers to $N$ distinct low energy
quanta rather than the excitation of a single high energy level. We
have found that the Born approximation to the amplitude attains a
maximum value of $\sim g^7\ 8! ,$ the expected behavior from the
analysis of tree graphs in a $g\phi^3$ theory. As a consequence,
unitarity breaks down for $\g2\simeq 0.07.$ For a range of $\g2$ near
this value, the holy grail function $F\simeq -0.20$ and shows only slow
$\g2$ dependence. This is in total contrast to the case of the anharmonic
oscillator. The amplitude is also about $200$ times greater than $e^{-N},\
N=8,$ another suggested form of exponential suppression. There is a
hint in this work that any exponential suppression of $\a1n$
which exists is
considerably weaker than that expected $(F\simeq -1)$ from Lipatov
type arguments. In another work, one of us [H.G.] will present an extension
of this model to arbitrary $N.$
\bigskip
\noindent{\us{\bf Acknowledgements}}. One of the authors [H.G.] would
like to express appreciation to the participants of the Yale-Texas
Workshop on Electroweak Baryon Number Violation for stimulating
discussion following a talk based on this work. The research of H.G. was
supported in part by the National Science Foundation under
Grant No. PHY-9001439, and by the Texas National Laboratory Research
Commission under Grant No. RGFY9114.
The research of M.T.V. was supported in part by the U.~S. Department
of Energy under Grant No.
DE-FG02-85ER40233.
\bigskip
\centerline{ FIGURE CAPTIONS}
\bigskip
\item{Fig.\ 1.} The time dependence of the probability for
transition from the 1-particle to the 8-particle state, for
$\g2=0.07.$ This is the value of the coupling at which the Born
approximation for $A_{1\ra 8}$ violates unitarity.
\item{Fig.\ 2.} Time dependence of the probability to stay in the
1-particle state, $\g2=0.07.$
\item{Fig.\ 3.} {\sl Solid line}: Maximum amplitude attained for reaching
the 8-particle state, as a function of $\g2.$ {\sl Dashed line:} Upper
bound on matrix element $\VEV{8|x|0}$ as a function of $\g2,$ from
ref.\ [\bachas].
\item{Fig.\ 4.} Same as Fig.\ 3, but for the holy grail function $F$
defined in Eq. \fdef .
\bigskip

\refout
\bye